\documentclass[fleqn,usenatbib]{mnras}

\usepackage[T1]{fontenc}

\DeclareRobustCommand{\VAN}[3]{#2}
\let\VANthebibliography\thebibliography
\def\thebibliography{\DeclareRobustCommand{\VAN}[3]{##3}\VANthebibliography}

\usepackage{graphicx}	
\usepackage{amsmath}	
\usepackage{amssymb}	
\usepackage{newtxtext,newtxmath}	
\usepackage{cleveref}	

\numberwithin{equation}{section}
\crefname{equation}{equation}{equation}
\crefname{figure}{Fig.}{Fig.}

\title[Neutron star oscillations]{Neutron star oscillations in pseudo-Newtonian gravity}

\author[Y.-T. Tang and L.-M. Lin]{
Yat-To Tang\thanks{Email address: yttang@phy.cuhk.edu.hk} and 
Lap-Ming Lin\thanks{Email address: lmlin@phy.cuhk.edu.hk} 
\\
Department of Physics, The Chinese University of Hong Kong, Hong Kong, China\\
}

\date{Accepted XXX. Received YYY; in original form ZZZ}

\pubyear{2021}

\begin{document}
\label{firstpage}
\pagerange{\pageref{firstpage}--\pageref{lastpage}}
\maketitle

\begin{abstract}
We investigate the oscillations of neutron stars using a purely Newtonian approach and three other 
pseudo-Newtonian formulations. 
Our work is motivated by the fact that pseudo-Newtonian formulations are commonly used in 
core-collapse supernova (CCSN) simulations.  
We derive and solve numerically the radial and nonradial perturbation equations for neutron star oscillations using different combinations of modified Newtonian hydrodynamics equations and gravitational potentials. We pay special 
attention to the formulation proposed recently by Zha et al. [Phys. Rev. Lett. 125, 051102 (2020)] that implements the standard Case A effective potential in CCSN simulations with an additional lapse-function correction to the hydrodynamics equations. We find that this ``Case A+lapse'' formulation can typically approximate the frequency of the fundamental radial mode of a $1.4 M_\odot$ neutron star computed in general relativity to about a few tens of percent for our chosen EOS models. 
For the nonradial quadrupolar $f$ mode, which is expected to contribute strongly to the gravitational waves emitted from a protoneutron star, the Case A+lapse formulation performs much better and can approximate the $f$ mode 
frequency to within about a few percent even for the maximum-mass configuration in general relativity. 
\end{abstract}




\section{Introduction}
\label{intro}

It is well known that the oscillation modes of a neutron star carry rich information about the internal structure of the star and the poorly understood nuclear-matter equation of state (EOS). 
Pulsating neutron stars are also potential sources of gravitational waves for current and future ground-based detectors. The detection of the gravitational-wave signals associated to neutron star oscillations can yield important information about the EOS, which would complement the constraints set by the measurement of the tidal 
deformability by the gravitational-wave signals of binary neutron stars 
\citep{GW170817}, and also the recent mass and radius measurements of pulsars 
in the electromagnetic channel \citep{Miller:2019,Riley:2019,Miller:2021,Riley:2021}.

The study of the oscillation modes of neutron stars in general relativity (GR) has a long history dating back to the seminal works of \citet{Chandra:1964} and \citet{Thorne:1967}. For nonrotating neutron stars, the computation of the oscillation modes are well established and can be formulated as eigenvalue problems \citep{Chandra:1964,Thorne:1967,Lindblom:1983,Detweiler:1985} by perturbing a spherically symmetric background solution. 
However, the situation for rapidly rotating neutron stars is more complicated, as the background solution
is no longer spherically symmetric and the stellar deformation due to rotation cannot be treated perturbatively at arbitrary spin frequencies.  
The oscillation modes of rapidly rotating neutron stars are currently best studied using a time-evolution approach under various approximations, such as the Cowling approximation 
\citep{Font:2001,Stergi:2004,Gaertig:2008}, the conformally flatness assumption \citep{dimmelmeier}, and in the linearized theory of GR \citep{Kruger:2020}. 
In this approach, one determines the oscillation modes by suitably perturbing and following the evolution of a rotating neutron star using either a linear or nonlinear hydrodynamics code. The oscillation modes
are identified and their frequencies are obtained by performing Fourier transforms for the matter variables
such as the fluid density and velocity fields (see, e.g., \citet{dimmelmeier}).

Hydrodynamical simulations may also be the only way to study the properties of the protoneutron star born in a core-collapse supernova (CCSN). 
Although the event rate is low, the gravitational-wave signals (if detected) of a fiercely pulsating protoneutron star within our galaxy can provide us a unique probe to the properties of the star and
the finite-temperature nuclear matter 
(e.g., \citet{Muller:2013,Sotani:2017,morozova,Torres:2019,Radice:2019,Bizouard:2021,Eggenberger:2021}). 
The signals may also encode information about a potential quantum chromodynamics (QCD) phase transition \citep{zha}.  
However, due to the high computational cost and the complexity of the problem, there are so far only 
limited attempts of CCSN simulations in full GR \citep{Liebend:2004,OConnor:2015,Kuroda:2016,Ott:2018} that take into account realistic microphysics inputs, such as neutrino transport, under various approximations. Relativistic simulations under the conformally flatness assumption have also been performed (e.g., \citet{Cerda:2013,Muller:2019}).  
In fact, many current state-of-the-art CCSN simulations have used either a purely Newtonian theory for gravity or a pseudo-Newtonian approach where phenomenological GR corrections are implemented in Newtonian hydrodynamics. These GR corrections include the so-called Case A effective potential 
proposed by \citet{marek}, which has been used extensively in CCSN simulations
\citep{scheidegger,yakunin,morozova,OConnor:2018,pan}. The Case A potential formalism has been
shown to be a good approximation to GR in simulating nonrotating or slowly rotating CCSN \citep{marek}. 
More recently, \citet{zha} have extended the Case A formalism by adding a lapse function to the 
Newtonian hydrodynamics equations to mimic the time-dilation effect (see also \citet{Obergaulinger:2020}).

The effectiveness of using the Case A potential formalism to approximate GR effects has been studied 
by comparing the results with relativistic simulations \citep{OConnor:2018,Pajkos:2019,Muller:2008}.
In particular, \citet{Muller:2008} found that the Case A potential formalism fails to capture the correct oscillation modes, and suggested the absence of a lapse function as the cause.   
\citet{zha} have also studied the performance of their extended formalism with a lapse function 
by simulating a radially oscillating protoneutron star (see the supplemental material of their paper). They found that adding the lapse function correction to the hydrodynamics equations can help to shift 
the fundamental radial mode frequency toward the GR result. 
In this paper, we aim to provide a systematic study of the radial and nonradial oscillation modes of
neutron stars in the Case A potential formalism with or without a lapse function correction. 
Instead of using dynamical simulations, we obtain the oscillation mode frequencies by 
perturbative calculations and compare them to the GR results. 
Being able to determine the mode frequencies from the perturbative equations that are consistent with 
CCSN simulations would be useful for identifying which oscillation modes are excited in the simulations \citep{WS2020}. 
Our study can also benchmark the accuracy of the mode frequencies obtained in pseudo-Newtonian simulations against exact GR results. 
This investigation is not only for theoretical interest, but should also be important to the detection of the gravitational wave signals from a nearby CCSN, as the ability to identify the mode frequencies of the protoneutron stars would help to tune the configurations of gravitational wave detectors to increase the sensitivity near those frequencies \citep{Srivastava:2019} and the chance of detection.

In this work, we derive the perturbation equations for radial and nonradial oscillations based on four different
perturbation schemes, including purely Newtonian one, in which different combinations of GR corrections for the unperturbed background star and perturbations are used. We then solve the different systems numerically to obtain the oscillation mode frequencies and compare the results with GR calculations. 
We find that including a lapse function correction to the Newtonian hydrodynamics 
equations generally can help to better approximate the GR mode frequencies. 
In particular, we find that using the Case A potential formulation with the lapse 
function correction as proposed by \citet{zha} can approximate the nonradial 
quadrupolar $f$ mode frequency to within about a few percent for the EOS and star models that we have studied. Interestingly, for a fixed central energy density, we also find that purely Newtonian calculation can also approximate the GR $f$ mode frequency to about
the same level of accuracy, though the Newtonian and GR background stars are quite different from each other. 

The plan of the paper is as follows. We introduce the four perturbation schemes and some general equations in Section~\ref{sec:method}. The perturbation equations and numerical schemes for radial and nonradial oscillations are presented in Sections~\ref{sec:radial} and \ref{sec:nonradial}, respectively. In Section~\ref{sec:test}, we present
various tests to validate our numerical codes. We then present our numerical results in 
Section~\ref{sec:results}. Finally, we conclude our paper in 
Section~\ref{sec:conclusions}. Unless otherwise noted, we use geometrical units with $G=c=1$.

\section{General equations}
\label{sec:method}


\begin{table*}
	\centering
	\caption{The schemes examined in this study and the corresponding background and perturbation equations
	to be solved for the oscillation modes. 
	Nonradial perturbation equations are the same for all four schemes, but some of them include a lapse-function $\alpha$ in the hydrodynamics equations. For schemes that do not include the lapse function, $\alpha$ is set to 1 and its derivative zero in the equations. Note that the lapse function only appears in the perturbation equations but not in the background equations.}
	\label{tab:schemes}
	\begin{tabular}{ccccc}
		\hline
		Scheme & Background equations & Radial perturbation equations &
		Nonradial perturbation equations & Lapse function $\alpha$ included?\\
		\hline
		N & Eqs.~(\ref{N_bg1}) to (\ref{N_bg3}) 
		 & Eqs.~(\ref{r_p1}), (\ref{r_p2}) and (\ref{r_dm_N}) 
		 & Eqs.~(\ref{n_p1}), (\ref{n_p2}), (\ref{n_p3}) and (\ref{n_p4}) & No\\
		N+lapse & Eqs.~(\ref{N_bg1}) to (\ref{N_bg3}) &
		 Eqs.~(\ref{r_p1}), (\ref{r_p2}) and (\ref{r_dm_N})  &
		 Eqs.~(\ref{n_p1}), (\ref{n_p2}), (\ref{n_p3}) and (\ref{n_p4})  & Yes\\
		Case A & Eqs.~(\ref{A_bg1}) to (\ref{A_bg3}) 
		& Eqs.~(\ref{r_p1}), (\ref{r_p2}) and (\ref{r_dm_A})  & 
		 Eqs.~(\ref{n_p1}), (\ref{n_p2}), (\ref{n_p3}) and (\ref{n_p4})  & No\\
		Case A+lapse & Eqs.~(\ref{A_bg1}) to (\ref{A_bg3}) &
		Eqs.~(\ref{r_p1}), (\ref{r_p2}) and (\ref{r_dm_A}) &
     	Eqs.~(\ref{n_p1}), (\ref{n_p2}), (\ref{n_p3}) and (\ref{n_p4}) 	  & Yes\\
		\hline
	\end{tabular}
\end{table*}

We study the oscillation modes of nonrotating fluid neutron stars, ignoring the effects of rotation 
and solid crust. 
Our perturbative mode calculation is performed by perturbing a spherically symmetric equilibrium background solution. It involves background equations (to be solved for the equilibrium background) and perturbation equations (to be solved for the oscillation modes). 
We study both the radial and nonradial oscillation modes based on four different perturbation schemes.
The four schemes include: purely Newtonian hydrodynamics (N), Newtonian hydrodynamics with lapse-function correction (N+lapse), Newtonian hydrodynamics with Case A potential (Case A), and Newtonian 
hydrodynamics with Case A potential and lapse-function correction (Case A+lapse). They differ from each other by the choices of the background and perturbation equations.

In this section, we first review the Case A potential formalism and present the relevant background equations and the general linearized hydrodynamics equations, based on which the radial and nonradial
oscillation mode equations will be derived. 
For convenience, we summarize the set of equations to be solved in each perturbation scheme in
\Cref{tab:schemes}. 
Readers who are more interested in the numerical results may refer to the table 
for a summary of our perturbation schemes and go directly to \Cref{sec:results}, skipping the 
derivations in Sections \ref{sec:radial} and \ref{sec:nonradial}.

\subsection{Review of the Case A potential}
\label{sec:caseA_review}

The Case A effective potential is defined by replacing the Newtonian gravitational potential in a 
spherically-symmetric Newtonian hydrodynamics simulation by \citep{marek}
\begin{equation}
    \Phi_{\rm TOV}(r)=-4\pi\int^\infty_r \frac{dr'}{r'^2} \left(\frac{m_{\rm TOV}}{4\pi}+r'^3 P\right) \times \frac{1}{\Gamma^2} \left(\frac{\rho+\rho\epsilon+P}{\rho}\right), \label{caseA_def1}
\end{equation}
where $r$ is the radial coordinate, $\rho$ is the rest-mass density, $P$ is the pressure, $\epsilon$ is the specific internal energy, and hence the total energy density is given by $e=\rho+\rho\epsilon$. 
The function $m_\text{TOV}$ is defined by
\begin{equation}
    m_{\rm TOV}(r) = 4\pi\int^r_0 dr' r'^2 \Gamma(\rho+\rho\epsilon) ,  \label{caseA_def2}
\end{equation}
where $\Gamma$ is given by
\begin{equation}
    \Gamma=\sqrt{1-2\frac{m_\text{TOV}}{r}}. \label{gamma_def}
\end{equation}
It should be noted that the local radial velocity (squared) is also included inside the square 
root of $\Gamma$ in the original definition for dynamical simulations \citep{marek}. However, 
we ignore that velocity term here as we only consider non-rotating background stars in our 
perturbation calculations. We have also ignored the contributions of neutrino energy and pressure introduced in the original definitions of $\Phi_\text{TOV}$ and $m_\text{TOV}$ for dynamical CCSN simulations \citep{marek}. 

Since we need to set up differential equations for perturbative calculations, 
equations~(\ref{caseA_def1}) and (\ref{caseA_def2}) are recast into the following differential forms:
\begin{align}
    &\frac{d\Phi_\text{TOV}}{dr}=\frac{4\pi}{r^2} \left(\frac{m_\text{TOV}}{4\pi}+r^3 P\right) \frac{1}{\Gamma^2} \left(\frac{\rho+\rho\epsilon+P}{\rho}\right)  , \label{caseA_diff1}\\
    &\frac{dm_\text{TOV}}{dr}=4\pi r^2 \Gamma (\rho+\rho\epsilon) . \label{caseA_diff2}
\end{align}
These equations will be employed in the background and radial perturbation equations for 
the Case A and Case A+lapse schemes.

\subsection{Equilibrium background equations}
\label{sec:bg}

The starting point of our perturbation calculation is the construction of an unperturbed equilibrium 
star by solving the background equations with a given EOS. Our four different perturbation schemes 
rely on different background equations. For the N and N+lapse schemes, the background equations are 
simply the standard Newtonian hydrostatic equilibrium equations:
\begin{align}
	&\frac{dm}{dr}=4\pi\rho r^2 , \label{N_bg1}\\
	&\frac{dP}{dr}=-\frac{\rho m}{r^2} , \label{N_bg2}\\
	&\frac{d\Phi}{dr}=-\frac{1}{\rho}\frac{dP}{dr}, \label{N_bg3}
\end{align}
where $\Phi$ is Newtonian gravitational potential and $m$ is the enclosed mass.

For the Case A and Case A+lapse schemes, the background equations are obtained by replacing the 
Newtonian gravitational potential by the Case A potential, and imposing the hydrostatic equilibrium 
equation~(\ref{N_bg3}):  
\begin{align}
	&\frac{dm}{dr}=4\pi(\rho+\rho\epsilon)r^2\Gamma \label{A_bg1}\\
	&\frac{dP}{dr}=-\frac{4\pi}{r^2}\left(\frac{m}{4\pi}+r^3 P\right)\frac{1}{\Gamma^2}(\rho+\rho\epsilon+P) \label{A_bg2}\\
	&\frac{d\Phi}{dr}=-\frac{1}{\rho}\frac{dP}{dr}, \label{A_bg3}
\end{align}
where equations~(\ref{A_bg1}) and (\ref{A_bg3}) are equivalent to equations~(\ref{caseA_diff1}) and (\ref{caseA_diff2}). These background equations closely resemble the Tolman-Oppenheimer-Volkov equations in exact GR. The factor of $\Gamma$ in equation~(\ref{A_bg1}), which does not appear in the original equation in GR, is introduced so that dynamical simulations using the Case A potential can better match GR simulations \citep{marek}.

To solve the background equations, we need to specify an EOS to relate $\rho$, $P$ and $\epsilon$. 
In this study, we use four different nuclear-matter EOS models 
that have been employed in the study of neutron star oscillations in GR \citep{andersson,kokkotas}. 
Using the same model names defined in these papers, we include their EOS A and B \citep{eosAB}, 
EOS C (which is model I in \citet{eosC}), and EOS F \citep{eosF}. 
We obtain the tabular data for these EOS models from an open-source 
code\footnote{http://www.gravity.phys.uwm.edu/rns/} {\tt RNS}.
It should be pointed out that these EOS models are outdated and have in fact been ruled out by 
the observations of neutron stars with masses $\approx 2 M_\odot$ \citep{Demorest:2010,Antoniadis:2013}. 
Nevertheless, we choose them because we can evaluate the performance of our pseudo-Newtonian
perturbation schemes by comparing our numerical results, both the radial and nonradial oscillation modes,  
with the exact GR results tabulated in \citep{andersson,kokkotas} systematically.

\subsection{Linearized fluid equations}
\label{sec:gne}

We start from the modified Newtonian hydrodynamics equations in \citet{zha}, where a lapse function $\alpha$
is added to mimic the time-dilation effect. The modified equations are 
\begin{align}
	&\frac{\partial\rho}{\partial t}+\nabla\cdot(\alpha\rho\vec{v})=0 , \label{Euler1_mod1}\\
	&\frac{\partial}{\partial t}(\rho\vec{v})+\nabla\cdot[\alpha(\rho\vec{v}\vec{v}+P)]=-\alpha(\rho-P)\nabla\Phi, \label{Euler2_mod1}
\end{align}
where $\vec{v}$ is the fluid velocity and $\Phi$ is the gravitational potential. The lapse function
is defined by $\alpha=\exp (\Phi)$. Equation~(\ref{Euler2_mod1}) can be rewritten as
\begin{equation}
	\rho\frac{\partial\vec{v}}{\partial t}+\alpha\rho\vec{v}\cdot\nabla\vec{v}=-\alpha(\nabla P+\rho\nabla\Phi). \label{Euler2_mod2}
\end{equation}

The hydrodynamics equations can be linearized by expanding the rest mass density, and similarly other physical 
quantities, by $\rho = \rho_0 + \delta\rho$, where $\rho_0$ is evaluated on the unperturbed background star
and $\delta$ denotes the Eulerian perturbation. Since the background star is non-rotating (i.e., 
${\vec v}_0 = 0$), the Lagrangian displacement of a fluid element is given by $\vec{\xi} = \delta {\vec v}$. 
The linearized hydrodynamics equations are given by 
\begin{align}
	&\delta\rho+\nabla\cdot(\alpha_0\rho_0\vec{\xi})=0 , \label{lin1_mod1}\\
	&\frac{\partial^2\vec{\xi}}{\partial t^2}=\alpha_0\left(\frac{\delta\rho}{\rho_0^2}\nabla P_0 - \frac{1}{\rho_0}\nabla\delta P-\nabla\delta\Phi\right). \label{lin2_mod1}
\end{align}

To determine the oscillation modes, we assume a time dependence of the form 
$\delta Q( {\vec r}, t) = \delta Q({\vec r})\exp(i \omega t)$ for all the perturbed variables and the 
Lagrangian displacement. 
The form of equation~(\ref{lin1_mod1}) remains unchanged after substituting 
this expansion, though it is understood that the perturbed variables $\delta\rho({\vec r})$ and 
${\vec \xi}({\vec r})$ are now functions of coordinates only. Similarly, equation~(\ref{lin2_mod1}) can be 
expressed as
\begin{equation}
	-\omega^2\vec{\xi}=\alpha_0\left(\frac{\delta\rho}{\rho_0^2}\nabla P_0 - \frac{1}{\rho_0}\nabla\delta P-\nabla\delta\Phi\right). \label{lin2_mod2}
\end{equation}

Equations~(\ref{lin1_mod1}) and (\ref{lin2_mod2}) are the main linearized fluid equations to be solved 
with the equation for the perturbed gravitational potential $\delta\Phi$ and the adiabatic condition (see
below) in order to derive different perturbation schemes. For the N and N+lapse schemes, $\delta\Phi$ is simply 
determined by the perturbed Newtonian Poisson equation
\begin{equation}
    \nabla^2\delta\Phi = 4\pi\delta\rho. \label{poisson}
\end{equation}
The situation for the Case A and Case A+lapse schemes will be discussed in the following section. For simplicity, the subscripts ``0'' for background quantities are dropped hereafter. That is, $\vec \xi$ and variables with 
$\delta$ represent perturbations, while variables without $\delta$, such as $\rho$ and $P$, simply denote the background quantities.

To end this section, we remark that the dynamical equations~(\ref{Euler1_mod1}) and (\ref{Euler2_mod2}), but not equation~(\ref{Euler2_mod1}), can reduce to the purely Newtonian equations in the case $\alpha=1$. 
Since the perturbed hydrodynamical equations presented in the following sections are derived from equations~(\ref{Euler1_mod1}) and (\ref{Euler2_mod2}), the Newtonian limit (i.e., the N scheme) can thus be recovered by simply setting $\alpha=1$ in our final sets of perturbation equations.

\section{Radial oscillations}
\label{sec:radial}
\subsection{Radial perturbation equations}
\label{sec:radial_osc}

To obtain the equations for radial oscillations, we assume spherical symmetry for the fluid motions, and hence the 
perturbed quantities $\delta Q ({\vec r}) = \delta Q(r)$ are functions of the radial coordinate $r$ only. 
Equations~(\ref{lin1_mod1}) and (\ref{lin2_mod2}) now become
\begin{align}
	&\delta\rho=-\left(\frac{d\alpha}{dr}\rho\xi+\alpha\frac{d\rho}{dr}\xi+\frac{2}{r}\alpha\rho\xi+\alpha\rho\frac{d\xi}{dr}\right) , \label{r_lin1_mod1}\\
	-&\omega^2\xi=\alpha\left(\frac{\delta\rho}{\rho^2}\frac{dP}{dr}-\frac{1}{\rho}\frac{d\delta P}{dr}-\frac{d\delta\Phi}{dr}\right), \label{r_lin2_mod1}
\end{align}
where $\xi$ is the radial component of $\vec{\xi}$, which equals to the magnitude of $\vec{\xi}$ due to spherical symmetry.

In this study, we consider adiabatic oscillations so that the perturbed fluid satisfies
\begin{equation}
\frac{\Delta P}{P} = \Gamma_1 \frac{\Delta \rho}{\rho},    
\label{eq:adiabatic}
\end{equation}
where $\Delta \rho$ and $\Delta P$ are the Lagrangian perturbations of density and pressure, respectively; 
$\Gamma_1$ is the adiabatic index for the perturbed fluid. In terms of the corresponding Eulerian perturbations, 
equation~(\ref{eq:adiabatic}) can be written as 
\begin{equation}
	\frac{\delta\rho}{\rho}=\frac{\delta P}{\Gamma_1 P} - A\xi, \label{r_adiabatic}
\end{equation}
where $A = \frac{1}{\rho}\frac{d\rho}{dr}-\frac{1}{\Gamma_1 P}\frac{dP}{dr}$ is the Schwarzschild discriminant.
By eliminating $\delta\rho$ using equation~(\ref{r_adiabatic}), equations~(\ref{r_lin1_mod1}) and (\ref{r_lin2_mod1}) now become
\begin{align}
	&\frac{d\xi}{dr}=\left(\frac{A}{\alpha}-\frac{1}{\alpha}\frac{d\alpha}{dr}-\frac{1}{\gamma P}\frac{dP}{dr}-\frac{2}{r}\right)\xi - \frac{1}{\alpha\Gamma_1 P}\delta P , \label{r_p1}\\
	&\frac{d\delta P}{dr}=\frac{\rho\omega^2}{\alpha}\xi + \left( \frac{\delta P}{\Gamma_1 P} - A\xi  \right)
	\frac{dP}{dr} -\rho\frac{d\delta\Phi}{dr} . \label{r_p2}
\end{align}
where $\gamma = d\ln{P}/d\ln{\rho}$ is defined for the unperturbed background star and the Schwarzschild 
discriminant can also be written as $A = \left(\frac{1}{\gamma} - \frac{1}{\Gamma_1} \right) \frac{d \ln P}{dr}$. 
Purely Newtonian perturbation equations (i.e., the N scheme) are obtained by setting the lapse function 
$\alpha = 1$. 

To close the perturbation equations, we need to obtain an equation for the perturbed gravitational potential 
$\delta \Phi$, which appears on the right hand side of equation~(\ref{r_p2}). 
For the N and N+lapse schemes in spherical symmetry, the perturbed Poisson equation~(\ref{poisson}) reduces to the following two first-order equations: 
\begin{align}
	&\frac{d\delta\Phi}{dr}=\frac{\delta m}{r^2}, \label{r_dphi_N} \\
	&\frac{d\delta m}{dr}=4\pi r^2 \delta\rho , \label{r_dm_N}
\end{align}
where $\delta\rho$ can be expressed in terms of $\delta P$ and $\xi$ by equation~(\ref{r_adiabatic}). 
In the N and N+lapse schemes, we substitute equation~(\ref{r_dphi_N}) into equation~(\ref{r_p2}) directly 
and solve equations~(\ref{r_p1}), (\ref{r_p2}) and (\ref{r_dm_N}) numerically for the oscillation modes. 

It should be pointed out that we can indeed make use of equation~(\ref{lin1_mod1}) to solve 
equation~(\ref{r_dm_N}) exactly and obtain $\delta m = - 4\pi r^2 \alpha\rho\xi$. 
Equations~(\ref{r_p1}) and (\ref{r_p2}) can then be combined to yield a second-order differential equation 
for $\xi$. We have checked that the resulting equation recovers the standard Newtonian equation for radial 
oscillations (see Table~\ref{tab:cox_compare}) when we set the lapse function $\alpha = 1$ (i.e., the N scheme). 
The reason that we formulate the problem by equations~(\ref{r_p1}), (\ref{r_p2}) and (\ref{r_dm_N}),
instead of a single second-order equation, is because we want to make a comparison to the Case A and Case A+lapse
schemes. 

For the Case A and Case A+lapse schemes, we perturb the Case A potential equation~(\ref{caseA_diff1}) and 
obtain
\begin{equation}
\begin{split}
	\frac{d\delta\Phi}{dr}=&\left(\frac{\delta m}{r^2}+4\pi r\delta P\right)
	\frac{1}{\Gamma^2}\left(1+\epsilon+\frac{P}{\rho}\right) \\
	&+\left(\frac{m}{r^2}+4\pi rP\right)\frac{1}{\Gamma^2}
	\left[\frac{2}{\Gamma^2}\frac{\delta m}{r}
	\left(1+\epsilon+\frac{P}{\rho}\right)
	+\frac{\delta P}{\rho}\right], \label{r_dphi_A}
\end{split}
\end{equation}
where we have used $\delta\epsilon=\frac{P}{\rho^2}\delta\rho$ and
$\delta\left(\frac{1}{\Gamma^2}\right)=\frac{2}{\Gamma^4}\frac{\delta m}{r}$.
Similarly, perturbing equation~(\ref{caseA_diff2}) gives
\begin{equation}
	\frac{d\delta m}{dr}=4\pi\Gamma\left(1+\epsilon+\frac{P}{\rho}\right)
	\delta\rho r^2 - 4\pi r\frac{1}{\Gamma}(\rho+\rho\epsilon)\delta m.
	\label{r_dm_A}
\end{equation}
As before, we do not solve equation~(\ref{r_dphi_A}) numerically, but substitute it directly into 
equation~(\ref{r_p2}). 

To sum up, for the N and N+lapse schemes, the radial perturbation equations are governed by 
equations~(\ref{r_p1})-(\ref{r_dm_N}). For the Case A and Case A+lapse schemes, 
equations~(\ref{r_p1}) and (\ref{r_p2}) are solved with equations~(\ref{r_dphi_A}) and (\ref{r_dm_A}). 
Furthermore, the lapse-function correction is turned off by setting $\alpha=1$ in the N and Case A schemes. 
As discussed in Section~\ref{sec:bg}, the four perturbation schemes are solved with different background equations. We summarize the background and perturbation equations for each scheme in Table~\ref{tab:schemes}.

\subsection{Boundary conditions and numerical scheme}
\label{sec:radial_num}

To solve the perturbation equations, we first need to obtain the unperturbed background solution. 
The calculation starts by specifying the initial condition $m(0)=0$ and a value for the central density $\rho(0)$, and hence $P(0)$ via the EOS, and integrate equations~(\ref{N_bg1}) and (\ref{N_bg2}) for the Newtonian
background (i.e., N and N+lapse schemes), or equations~(\ref{A_bg1}) and (\ref{A_bg2}) for the Case A background
(i.e., Case A and Case A+lapse schemes), outward to the stellar surface, where the radius $R$ is defined by the condition $P(R)=0$. The gravitational potential is then determined by solving equation~(\ref{N_bg3}) (or
equivalently equation~(\ref{A_bg3})) by imposing a suitable boundary condition at the surface. 
For the Newtonian background, it is given by
\begin{equation}
	\Phi(R) = -\frac{M}{R} .
\end{equation}
For the Case A background, we require
\begin{equation}
	\Phi(R) = \frac{1}{2}\ln{ \left(1-2\frac{M}{R} \right)}.
\end{equation}

With the Newtonian or Case A background solution determined, we then determine $\omega^2$ as an eigenvalue problem by solving the corresponding perturbation equations using a shooting method. 
The different sets of perturbation equations have the same boundary conditions $\xi(0)=\delta m(0)=0$ at 
the center. The value $\delta P(0)$ can be chosen arbitrarily. It is also recalled that the differential  
equation for $\delta \Phi$ (i.e., equation~(\ref{r_dphi_N}) or (\ref{r_dphi_A})) is not solved numerically,
but is substituted directly into the $d\delta P/dr$ equation, and hence there is no need to 
specify $\delta \Phi(0)$. 
Like the standard treatment of oscillation-mode calculations, the boundary condition at the surface 
is the vanishing of the Lagrangian perturbation of the pressure, $\Delta P(R)=0$, which can be expressed as 
\begin{equation}
	\delta P+\frac{dP}{dr}\xi=0 . 
\end{equation}
The values of $\omega^2$ that lead to the fulfillment of this boundary condition is the desired oscillation mode 
solutions.

\section{Nonradial oscillations}
\label{sec:nonradial}
\subsection{Nonradial perturbation equations}
\label{sec:nonradial_osc}

For nonradial oscillations, we will focus on the quadrupolar $f$-mode (fundamental) and $p$-modes (pressure) which belong to the class of spheroidal modes \citep{McDermott:1988}. The perturbation of scalar fields are expanded in spherical harmonics and the Lagrangian displacement is expanded in vector spherical harmonics:
\begin{align}
	&\delta\rho = \delta\tilde{\rho}(r)Y_{lm} , \\
	&\delta P = \delta\tilde{P}(r)Y_{lm} , \\
	&\delta\Phi =  \delta\tilde{\Phi}(r)Y_{lm} , \\
	&\vec{\xi} = U(r) Y_{lm} {\hat r}  +V(r) \nabla Y_{lm}  , 
\end{align}
where $Y_{lm} (\theta, \phi)$ is the standard spherical harmonics and $\hat r$ is the radial unit vector. 
Substituting these expansions into equations~(\ref{lin1_mod1}) and (\ref{lin2_mod2}) yields the following 
perturbation equations: 
\begin{align}
	&\frac{dU}{dr}=-\left(\frac{2}{r}+\frac{1}{\alpha}\frac{d\alpha}{dr}
	+ \frac{1}{\gamma P}\frac{dP}{dr}\right)U
	- \frac{1}{\alpha\rho}\delta\tilde{\rho}
	+ \frac{l(l+1)}{r}V , \label{n_p1}\\
	&\frac{d\delta\tilde{P}}{dr}=\frac{\rho\omega^2}{\alpha}U
	+ \frac{1}{\rho}\frac{dP}{dr}\delta\tilde{\rho}
	- \rho\frac{d\delta\tilde{\Phi}}{dr} , \label{n_p2}\\
	&V = \frac{\alpha}{r\omega^2}\left(\frac{1}{\rho}\delta\tilde{P}
	+ \delta\tilde{\Phi}\right) , \label{n_iv1}
\end{align}
where the first equation comes from equation~(\ref{lin1_mod1}), while the remaining two come from 
equation~(\ref{lin2_mod2}).
As before, $\delta\tilde{\rho}$ can be expressed in terms of other variables by the adiabatic condition. Expanded in spherical harmonics, the adiabatic condition (\ref{eq:adiabatic}) becomes
\begin{equation}
	\delta\tilde{\rho}=\frac{\rho}{\Gamma_1 P}\delta\tilde{P}-\rho AU  . 
	\label{n_iv2}
\end{equation}

In contrast to the case of radial oscillations, the perturbed Newtonian Poisson equation~(\ref{poisson}) will be used in all four nonradial perturbation schemes. This is because the Case A effective potential is a monopole potential, which does not have nonradial contributions. More precisely, the Case A potential formalism for multi-dimensional 
flow is constructed by defining the following effective potential \citep{marek}: 
\begin{equation}
	\Phi_\text{eff}(r,\theta,\phi) = \Phi(r,\theta,\phi) - \bar{\Phi}(r) + \bar{\Phi}_\text{TOV}(r),
\end{equation}
where $\Phi(r,\theta,\phi)$ is the standard Newtonian potential described by the Poisson equation, $\bar{\Phi}(r)$ is the radial part of the Newtonian potential, and $\bar{\Phi}_\text{TOV}(r)$ is the monopole potential given by equation~(\ref{caseA_def1}). Since $\bar{\Phi}(r)$ and $\bar{\Phi}_\text{TOV}(r)$ are radial functions only, when expanded in spherical harmonics, these terms have no contributions for $l \neq 0$. 

The perturbed potential $\delta \Phi$ for nonradial oscillations is thus simply described by the perturbed Poisson equation~(\ref{poisson}). After separating the angular parts, the radial function $\delta{\tilde \Phi}$ is determined by 
\begin{align}
	&\frac{d\Psi}{dr}=-\frac{2}{r}\Psi+\frac{l(l+1)}{r^2}\delta\tilde{\Phi}
	+4\pi\delta\tilde{\rho} , \label{n_p3}\\
	&\frac{d\delta\tilde{\Phi}}{dr}=\Psi, \label{n_p4}
\end{align}
where we have defined a new variable $\Psi$. As a result, we obtain four perturbation equations~(\ref{n_p1}), 
(\ref{n_p2}), (\ref{n_p3}), and (\ref{n_p4}), with the intermediate variables $V$ and $\delta\rho$ given by 
equations~(\ref{n_iv1}) and (\ref{n_iv2}), respectively. 
Thus, the N+lapse and Case A+lapse schemes are governed by the same set of nonradial perturbation equations, though
their background solutions are different. The N and Case A perturbation systems are obtained by setting the
lapse function $\alpha=1$ in the equations. 
We summarize the nonradial perturbation equations for each scheme in 
Table~\ref{tab:schemes}.

\subsection{Boundary conditions and numerical scheme}
\label{sec:nonradial_bc}

To solve the four differential equations~(\ref{n_p1}), (\ref{n_p2}), (\ref{n_p3}) and (\ref{n_p4})
for nonradial oscillation modes, boundary conditions at the center and surface are required. 
First the regularity conditions of the variables at the center yield the following relations:
\begin{align}
    & U = r^{l-1}A_0  , \\
	& \delta\tilde{P} = r^l B_0 , \\
	& \delta\tilde{\Phi} = r^l C_0  , \\
	& \Psi = lr^{l-1} C_0 , \\
	& A_0 = \frac{\alpha l}{\rho\omega^2} (B_0 + \rho C_0) ,
\end{align}
where $B_0$ and $C_0$ are constants. The surface boundary conditions are
\begin{align}
	\frac{dP}{dr}U+\delta\tilde{P}=0 ,  \label{eq:surface_BC1} \\
	\Psi=-\frac{l+1}{r}\delta\tilde{\Phi},  \label{eq:surface_BC2} 
\end{align}
where the first equation comes from requiring the Lagrangian perturbation of pressure to vanish at the surface, and the second comes from the continuity of $\delta\tilde{\Phi}$ and $d\delta\tilde{\Phi}/dr$.
The derivation of the above conditions is presented in \Cref{sec:bc}. 
It is noted that the lapse function correction is turned off (i.e., $\alpha=1$) in the above equations 
for the N and Case A perturbation schemes.

The nonradial perturbation equations~(\ref{n_p1}), (\ref{n_p2}), (\ref{n_p3}) and (\ref{n_p4}) can be written
formally as the matrix equation 
\begin{equation}
    \frac{d {\bf Y}}{dr} = {\bf Q} \cdot {\bf Y} , 
\end{equation}
where ${\bf Y} = (U, \delta {\tilde P}, \delta {\tilde \Phi}, \Psi)$ is an abstract vector formed by the perturbed variables. The matrix $\bf Q$ depends on $l$, $\omega^2$ and the background solution. 
Once an unperturbed background solution is obtained, the perturbation equations are solved by first choosing a value for $\omega^2$ and integrating the equations from the center. At $r=0$, we choose two orthogonal sets of constants $(A_0 , C_0)$, such as $(0,1)$ and $(1,0)$. With these choices, two sets of perturbation variables $(U(0),\delta\tilde{P}(0),\delta\tilde{\Phi}(0),\Psi(0))$ are obtained. 
The perturbation equations are integrated to $r=R/2$ for each set of initial variables, and two linearly 
independent solutions ${\bf Y}_1 (r)$ and ${\bf Y}_2 (r)$ are obtained. The general solution in the domain 
$0 \leq r \leq R/2$ is then given by a linear combination of the form $k_1 {\bf Y}_1 (r) + k_2 {\bf Y}_2 (r)$, where $k_1$ and $k_2$ are constants to be determined. 

The second part of the integration starts from the surface by choosing two sets of initial perturbed variables, 
${\bf Y}_3 (R)$ and ${\bf Y}_4(R)$, that satisfy the surface boundary conditions (\ref{eq:surface_BC1}) and (\ref{eq:surface_BC2}). The perturbation equations are then integrated backward from $R$ to $R/2$ with each set of initial variables, and hence a general solution $k_3 {\bf Y}_3(r) + k_4 {\bf Y}_4 (r)$ is obtained in 
the domain $R/2 \leq r \leq R$. At $r=R/2$, the two general solutions should equal to each other:
\begin{equation}
	k_1 {\bf Y}_1 (R/2)+ k_2 {\bf Y}_2 (R/2) = k_3 {\bf Y}_3 (R/2) + {\bf Y}_4 (R/2) , 
\end{equation}
where we have chosen the constant $k_4 = 1$ since it amounts to just an arbitrary normalization. The above equation can be rewritten as the following matrix equation
\begin{equation}
	\begin{pmatrix}
		Y_{1,1} & Y_{2,1} & Y_{3,1} \\
		Y_{1,2} & Y_{2,2} & Y_{3,2} \\
		Y_{1,3} & Y_{2,3} & Y_{3,3}
	\end{pmatrix}
	\begin{pmatrix}
		k_1 \\ k_2 \\ -k_3
	\end{pmatrix}
	=
	\begin{pmatrix}
		Y_{4,1} \\ Y_{4,2} \\ Y_{4,3}
	\end{pmatrix}
	, \label{n_matrix1}
\end{equation}
together with an algebraic equation
\begin{equation}
	k_1 Y_{1,4} + k_2 Y_{2,4} - k_3 Y_{3,4} - Y_{4,4}=0, \label{n_matrix2}
\end{equation}
where $Y_{i,j}$ is the $j$ component of ${\bf Y}_i (R/2)$. For a given value of $\omega^2$ and a background 
solution, equation~(\ref{n_matrix1}) is used to determine the constants $k_1$, $k_2$ and $k_3$. 
The value of $\omega^2$ is the desired oscillation mode solution if equation~(\ref{n_matrix2}) is satisfied. 
Otherwise, a new guess for $\omega^2$ is prescribed and the above procedure is repeated until an oscillation
mode solution is found to satisfy both equations~(\ref{n_matrix1}) and (\ref{n_matrix2}).

\section{Code Test}
\label{sec:test}

Before comparing the performance of our perturbation schemes to approximate the mode frequencies of neutron stars in GR, we first present various tests to validate our perturbation equations and numerical 
codes. In the tests, the two adiabatic indices $\gamma$ and $\Gamma_1$ are equal to each other, 
which is the same assumption that we used to obtain the numerical results in~\Cref{sec:results}.

To check that our codes can correctly reproduce the Newtonian results using the N scheme, we computed the radial and $l=2$ quadrupolar mode frequencies of a Newtonian polytropic star with polytropic index $n=1.5$, and compared our results with the data given in Table 17.2c of \citet{Cox}. 
We consider the fundamental radial $F$ mode and its first two overtones $H_1$ and $H_2$, and also the 
fundamental quadrupolar $f$ mode and the first two pressure modes $p_1$ and $p_2$. The mode frequencies and percentage differences between our results and those given by~\citet{Cox} are tabulated in Table~\ref{tab:cox_compare}. It is seen that our results agree very well with \cite{Cox} to within about
0.1\% accuracy.

\begin{table}
    \centering
    \caption{Comparison between the mode frequencies computed in our N scheme ($\Omega^2$) and those in \citet{Cox} ($\Omega^2_{\text{cox}}$) for the same Newtonian polytropic star with polytropic index $n=1.5$. All frequencies are expressed in dimensionless form ($\Omega^2=\omega^2 R^3/M$). }
    \begin{tabular}{c c c c}
        \hline
        Mode & $\Omega^2_{\text{cox}}$ & $\Omega^2$ & \% Difference \\
        \hline
        $F$ & 2.706 & 2.698 & 0.3\% \\
        $H_1$ & 12.54 & 12.53 & 0.1\% \\
        $H_2$ & 26.58 & 26.57 & 0.0\% \\
        $f$ & 2.119 & 2.123 & 0.2\% \\
        $p_1$ & 10.29 & 10.27 & 0.2\% \\
        $p_2$ & 23.52 & 23.51 & 0.1\% \\
        \hline
    \end{tabular}
    \label{tab:cox_compare}
\end{table}

We next turn to the Case A scheme where the effective potential defined originally by equation~(\ref{caseA_def1})
is only designed to mimic relativistic effects and is not based on a fundamental equation of motion, such as
the Poisson equation in Newtonian gravity. In our perturbation scheme, we recast the effective potential into a differential equation~(\ref{caseA_diff1}) and perturb it on an ad hoc basis. 
To check that our formulation gives the correct solution, we benchmark the mode frequencies computed in our Case A scheme against the results obtained in pseudo-Newtonian hydrodynamic simulations.

In Table~\ref{tab:WSMuller_compare}, we compare our Case A results with the simulation data given in \citet{Muller:2008} and \citet{WS2018}. Both of these works computed the oscillation modes of the same polytropic star model 
by perturbing and evolving the star hydrodynamically in the Case A formulation. The star model is described by a polytropic index $n=1$, polytropic constant $k=1.4553\times10^5$ g$^{-1}$cm$^5$s$^{-2}$, and central density $\rho_c=7.9\times10^{14}$ gcm$^{-3}$ 
(see Section 4.4 in \citet{Muller:2008} or Table 16.3 in \citet{WS2018}). 
Since \citet{Muller:2008} did not provide the exact numerical values, the frequencies reported in Table~\ref{tab:WSMuller_compare} are extracted directly from their figure. 
We find that the fundamental radial $F$ mode frequency computed in our Case A scheme agrees to both simulation results very well to within 1\%. 
We also notice that the $H_1$ and $H_2$ mode frequencies obtained by the two simulations differ 
significantly. Nevertheless, we find that our perturbative results for the radial and nonradial modes agree very well with the simulation results of \citet{WS2018}.

\begin{table}
    \centering
    \caption{Comparison between the radial and nonradial mode frequencies (in Hz) of a polytropic star model
    (see text for details) computed using the Case A scheme in our perturbative calculations and the simulations performed by \citet{Muller:2008} and \citet{WS2018}. Note that the nonradial mode frequencies for this star model are not provided by \citet{Muller:2008}.     }
    \begin{tabular}{c c c c}
        \hline
        Mode & M\"uller et al. &  Westernacher-Schneider & Our work \\
        \hline
        $F$ & 2170 & 2174 & 2157\\
        $H_1$ & 4350 & 5522 & 5529\\
        $H_2$ & 6540 & 8295 & 8290\\
        $f$ & -- &  2024 & 2021\\
        $p_1$ & -- &  5122 & 5131\\
        $p_2$ & -- & 7920 & 7932\\
        $p_3$ & -- &  10593 & 10603\\
        \hline
    \end{tabular}
    \label{tab:WSMuller_compare}
\end{table}

\citet{Muller:2008} also studied the oscillation modes of the same star model in Table~\ref{tab:WSMuller_compare}
by performing GR simulation under the conformal flatness approximation. The obtained $F$ mode frequency is 
1530 Hz, which is about $45\%$ lower than the result of Case A simulation (2170 Hz). They suggested that 
the large error may be due to the missing of a lapse function in the Case A simulation. To check this hypothesis, 
we studied the effect of a lapse function by computing the same star model using the Case A+lapse scheme. We obtained an $F$ mode frequency of 1699 Hz, which is only 11\% away from their GR result. This confirms their suggestion that the lapse function plays a significant role in approximating the mode frequencies in pseudo-Newtonian simulations.

As \citet{zha} performed simulations with both the Case A potential and a lapse function correction
(see the supplementary material in their paper), we also compare our Case A+lapse scheme with their results. 
They simulated a protostar model that contains a quark matter core using the FLASH code \citep{Flash} in two scenarios, one with the Case A potential with a lapse function correction, the other with the Case A potential 
only. In the former case, they reported the $F$ mode frequency at about 3300 Hz, while the frequency increases to about 4500 Hz in the latter. As a comparison, our perturbative calculations give 3096 Hz and 3915 Hz respectively for the two cases, showing a reasonable agreement with their simulation results. 
However, the percentage difference ($13\%$) between our $F$ mode frequency in the Case A scheme and their simulation result is much larger than the differences ($< 1\%$) for the $F$ mode of a
polytropic star that we compare with the simulations of \citet{Muller:2008} and \citet{WS2018} in 
Table~\ref{tab:WSMuller_compare}. It is unclear to us what causes the larger difference between our results and the simulations of \citet{zha}.

\section{Numerical results}
\label{sec:results}

In this section, we calculate the oscillation modes and compare the different perturbation schemes in this study using the EOS models A, B, C, and F employed in \citet{andersson} and \citet{kokkotas}. 
We assume barotropic oscillations so that the background star and the perturbed fluid are 
described by the same EOS. The two adiabatic indices $\gamma$ and $\Gamma_1$ are then equal and the 
Schwarzschild discriminant $A=0$. 

In the N and N+lapse schemes, only the rest-mass density $\rho$ appears in the 
background and perturbation equations, but not the total energy density $e=\rho+\rho\epsilon$. However, both 
$\rho$ and $e$ appear explicitly in the Case A and Case A+lapse schemes, and hence this poses an ambiguity 
on the comparison of different schemes. For instance, should we compare the results for a star model constructed
with a given central rest-mass density $\rho_c$ or energy density $e_c$? 
As \citet{andersson} and \citet{kokkotas} tabulated their results using the central energy density $e_c$, 
we shall also use $e_c$ as the parameter for comparing our different schemes with their GR results. 
In the N and N+lapse schemes, for a given central energy density $e_c$, the central rest-mass density $\rho_c$ 
that is needed as an initial condition for constructing the background star can be obtained through multiplying the number density given in the EOS data by baryon mass, which we have taken to be the same as the proton mass 
$m_p = 1.66\times10^{-24}$ g.

In the following, we compute the equilibrium background solutions, radial modes and quadrupolar ($l=2$) nonradial
modes of various star models for each perturbation scheme, and compare with the GR results in \citep{andersson, kokkotas}.

\subsection{Equilibrium background}
\label{sec:results_bg}

\begin{figure}
	\includegraphics[width=\columnwidth]{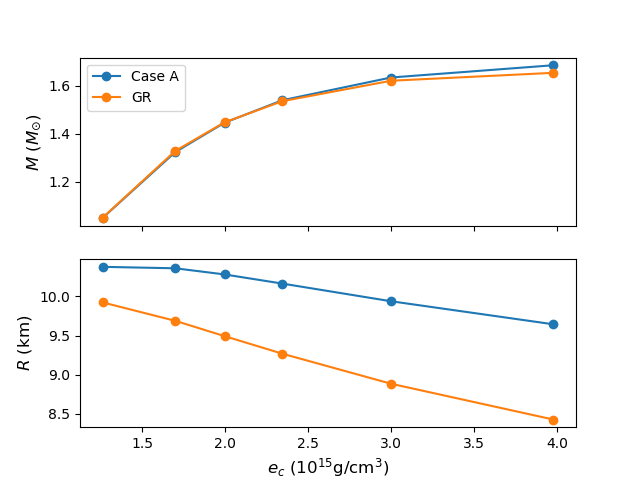}
    \caption{Mass and radius of stellar models for EOS A, as a function of central energy density $e_c$. 
    The Case A lines represent the background stars calculated by the Case A effective potential formulation
    (see Table~\ref{tab:schemes}). The GR lines are the exact GR solutions computed by \citet{andersson}.  }
    \label{fig:bg_eosA}
\end{figure}

As an illustration, Fig.~\ref{fig:bg_eosA} shows the mass and radius of star models for EOS A, plotted as a function of central energy density $e_c$. The Case A lines represent the results obtained by the Case A effective potential formulation where equations~(\ref{A_bg1}) to (\ref{A_bg3}) are used. The exact GR results of \citet{andersson} are given by the GR lines. The results for EOS B, C, and F are qualitatively the same as those for EOS A, and thus they are not shown. 

In general, the masses computed by the Case A formulation can approximate the GR results very well. The 
percentage differences between the two results for our EOS models are about 1\% for a $1.4 M_\odot$
star model, though the differences increase to about 3\% for the maximum-mass configurations. 
On the other hand, depending on the value of $e_c$ and the EOS model, the percentage difference of the stellar radii computed by the Case A formulation and GR ranges from about 5\% to 17\% (see the tables in 
Appendix \ref{sec:tabulated}). The Newtonian results are not plotted in the figure as they deviate significantly
from the GR results as expected for neutron star models. 
For instance, the Newtonian background star can reach up to $M=15 M_\odot$ at $e_c \approx 4\times 10^{15} \
{\rm g cm}^{-3}$ and the radius is only about 18 km for EOS A, causing the star to have a compactness $M/R > 1$.

\begin{figure}
	\includegraphics[width=\columnwidth]{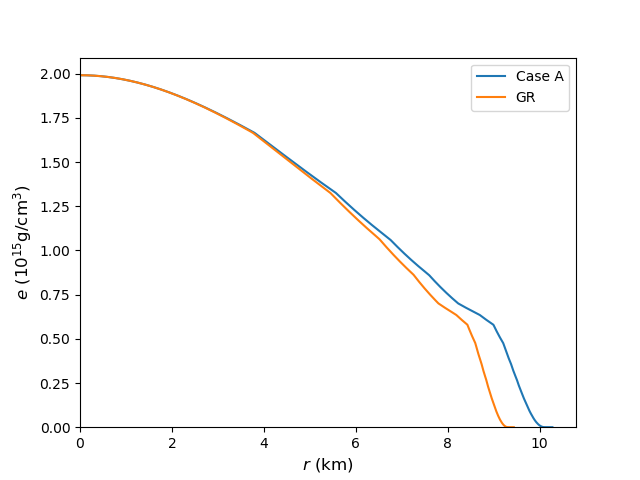}
	\caption{Comparison of the energy density profiles of the GR and Case A background solutions for a star
	 model governed by EOS A.}
	\label{fig:bg_profile}
\end{figure}

In Fig.~\ref{fig:bg_profile}, we compare the energy density profiles computed by the Case A and GR formulations, for a star model of central energy density $e_c = 2.0\times 10^{15}\ {\rm g cm}^{-3}$ governed by EOS A. 
The figure shows that the Case A solution has a noticeable deviation from the GR solution only in the outer region near the surface. As the total mass is contributed mainly by the high-density region of the 
star, this explains why the total mass computed in the Case A formulation agrees very well with that of the GR
calculation, though the radius has a larger deviation.

\subsection{Radial oscillations}
\label{sec:results_r}

\begin{figure}
	\includegraphics[width=\columnwidth]{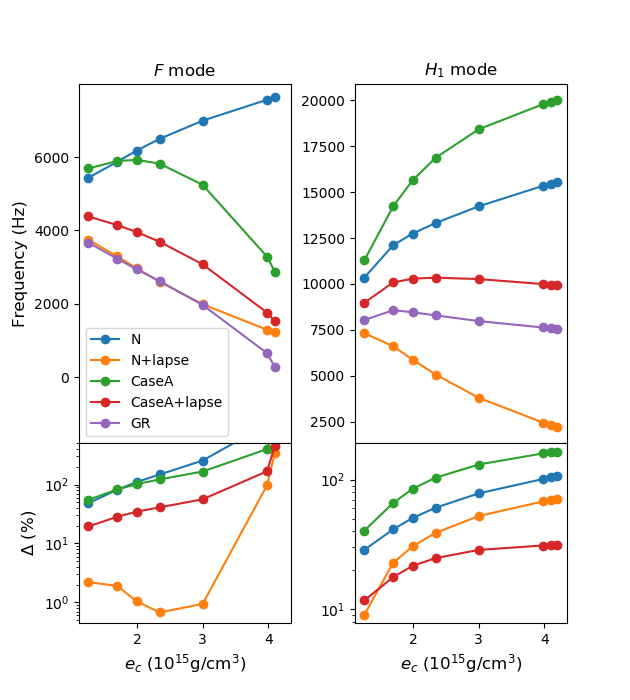}
    \caption{The upper panels plot the frequencies of the fundamental ($F$) radial mode and its first overtone ($H_1$) for EOS A against the central energy density $e_c$ for each of the 4 perturbation schemes. 
    The GR curves correspond to the GR results calculated by \citet{kokkotas}. 
    The lower panels show the percentage difference $\Delta$ defined by equation~(\ref{delta}) between our 
    numerical results and the GR results. }
    \label{fig:r_eosA}
\end{figure}

After studying the properties of the background solutions, we now compare the radial oscillation modes obtained by the four perturbation schemes (see Table~\ref{tab:schemes}) with those calculated by GR. 
Fig.~\ref{fig:r_eosA} shows the frequencies of the fundamental radial mode ($F$) and its first overtone ($H_1$)
as a function of central energy density $e_c$ for EOS A in the upper left and right panels, respectively. 
In the figure, our results for the N, N+lapse, Case A, and Case A+lapse perturbation schemes are compared with 
the GR results of \citet{kokkotas}. As noted above, in the N and N+lapse schemes, the corresponding central rest-mass density $\rho_c$ that is needed as an initial condition for the calculation is determined by $e_c$ through the EOS. 
The lower panels in the figure show the absolute percentage difference $\Delta$ defined by
\begin{equation}
	\Delta = \left\lvert\frac{f-f_\text{GR}}{f_\text{GR}}\right\rvert \times 100\%, \label{delta}
\end{equation}
where $f$ is the mode frequency computed by our perturbation schemes and $f_\text{GR}$ is the corresponding frequency in the GR solution.
In general, all the perturbation schemes deviate more from GR as the central density increases. For the $F$ mode, the percentage differences increase rapidly, especially at higher density, to more than 100\%. This is because the $F$ mode frequency in GR decreases towards zero as the background star gets closer to the maximum mass 
configuration, amplifying the percentage differences.  

The $F$ mode frequency in the purely Newtonian (N) calculation simply increases with the density and deviates largely from the GR frequency as expected. However, the other three schemes with relativistic corrections can capture qualitatively the trend of the GR results, namely the decrease of the $F$ mode frequency as $e_c$ increases. 
The inclusion of the lapse function correction in the hydrodynamics equations also has a drastic effect on both
the $F$ and $H_1$ mode frequencies. 
Interestingly, we find that the N+lapse scheme shows a better approximation for the $F$ mode, comparing to 
the Case A and Case A+lapse schemes, though the Newtonian background solution deviates a lot from the GR solution. 
This is generally true for the other EOS models that we have considered, but the percentage difference $\Delta$ is sensitive to the EOS. 
For EOS A, the N+lapse $F$ mode frequency agrees with the GR solution to within 2$\%$ level at low density, but the deviation can increase to about 10$\%$ for other EOS models at low density.

\begin{figure}
	\includegraphics[width=\columnwidth]{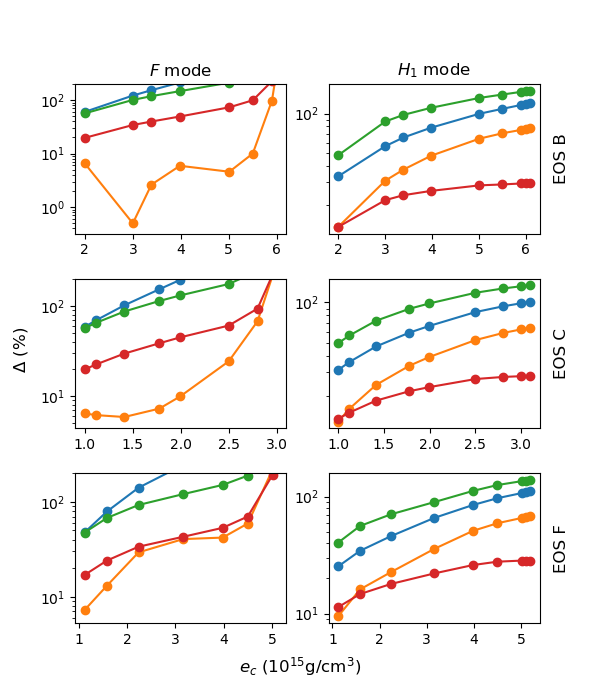}
	\caption{ Similar to the lower panels of Fig.~\ref{fig:r_eosA}, but for EOS models B, C, and F.}
	\label{fig:residue_radial}
\end{figure}

In Fig.~\ref{fig:residue_radial}, we plot the results of $\Delta$ for EOS B, C, and F. 
The color lines follow the same meanings as in Fig.~\ref{fig:r_eosA}. For instance, the red lines represent 
the results for the Case A+lapse scheme. 
While the N+lapse scheme can approximate the $F$ mode frequency to within about 10\% for some EOS models at 
low density, its deviation from the GR calculation increases quite rapidly with $e_c$ for the $H_1$ mode. 
The Case A+lapse scheme, on the other hand, can do a slightly better job and approximate the $H_1$ mode frequency 
to within about 20\% to 30\% in general. 
We tabulate the results of the $F$ and $H_1$ mode frequencies computed by the 
Case A+lapse scheme in Appendix~\ref{sec:tabulated} for reference. 
We also see that the N and Case A schemes do not show good approximations for the $F$ and $H_1$ mode frequencies. Nevertheless, including a lapse function correction in these schemes can help to bring their results closer to the GR mode frequencies.

\subsection{Nonradial oscillations}
\label{sec:results_n}

\begin{figure}
	\includegraphics[width=\columnwidth]{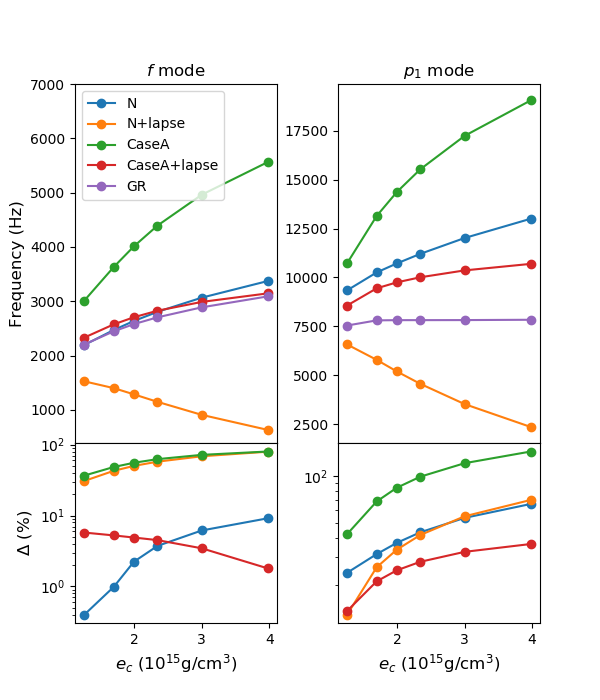}
    \caption{The upper panels plot the frequencies of the nonradial $f$ and $p_1$ modes for EOS A against the
    central energy density $e_c$ for each of the 4 perturbation schemes. The GR curves correspond to the GR
    results calculated by \citet{andersson}. The lower panels show the percentage difference $\Delta$ between
    our numerical results and the GR results. }
    \label{fig:n_eosA}
\end{figure}

\begin{figure}
	\includegraphics[width=\columnwidth]{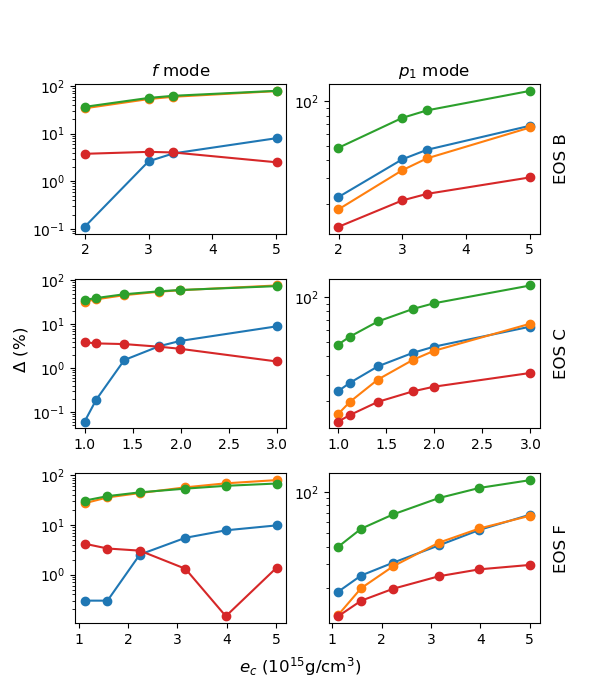}
	\caption{Similar to the lower panels of Fig.~\ref{fig:n_eosA}, but for EOS models B, C, and F.  }
	\label{fig:n_eosBCF}
\end{figure}

Having discussed the properties of the radial oscillations, we now turn our focus to the nonradial oscillation modes. In particular, we shall consider the $l=2$ quadrupolar fundamental ($f$) and first pressure ($p_1$) modes. Fig.~\ref{fig:n_eosA} plots the frequencies of the $f$ mode (left upper panel) and $p_1$ mode (right upper panel) against the central energy density for EOS A. As before, the results computed by the four different perturbation schemes are represented by different color lines in the figure. The GR curves in the upper panels represent the GR results obtained by \citet{andersson}. The lower panels show the absolute percentage differences $\Delta$ defined by equation~(\ref{delta}). 

While we have seen that the N+lapse scheme can approximate the radial $F$ mode reasonably well, 
Fig.~\ref{fig:n_eosA} shows that the nonradial $f$ and $p_1$ modes computed by this scheme deviate a lot 
from the GR results. In particular, the $f$ mode frequency of this scheme decreases with increasing $e_c$, which
is opposite to the trend of the GR solution and also the other schemes. On the other hand, the Case A+lapse scheme can give a good approximation to the $f$ mode frequencies to within a few percent level. 
Its percentage difference decreases to less than 2\% as $e_c$ rises to about
$4\times 10^{15}\ {\rm gcm}^{-3}$, the density at which the GR background star is close to the maximum-mass limit. 
For the $p_1$ mode, the performance of the Case A+lapse scheme is not as good as that for the $f$ mode, but it is still the best among the four perturbation schemes despite its percentage difference ranging from 10$\%$ to 
30$\%$. 

Fig.~\ref{fig:n_eosA} also shows, somewhat surprisingly, that the purely Newtonian N scheme can also provide a good approximation to the $f$ mode frequency, though the Newtonian and GR background stars with the same $e_c$ are quite different from each other. We recall that the central rest-mass density $\rho_c$ corresponding to a given 
$e_c$ that is needed for the Newtonian calculation is obtained via the EOS. 
The percentage difference of the N scheme is even smaller than that of the Case A+lapse scheme at low density. It
then increases to about 10\% at $e_c \approx 4\times 10^{15} \ {\rm gcm}^{-3}$, where the Case A+lapse scheme
becomes a better approximation. However, the N scheme does not perform as good as the Case A+lapse scheme for the $p_1$ mode across the whole density range in Fig.~\ref{fig:n_eosA}. 

We also plot the results of $\Delta$ for EOS B, C, and F in Fig.~\ref{fig:n_eosBCF} for comparison. 
The left and right panels in the figure correspond to the results for the $f$ and $p_1$ mode frequencies, respectively. 
The general trend that we have seen for EOS A in Fig.~\ref{fig:n_eosA} is also true for the other EOS models. 
In particular, we find that the N and Case A+lapse schemes can approximate the GR $f$ mode frequency very well, 
to within 0.1\% to 10\%, depending on the scheme and the value of $e_c$. 
If one takes into account the accuracy of the background stars, then the Case A+lapse scheme is seen to provide 
the best approximation to the $f$ mode frequency. We refer the reader to Appendix~\ref{sec:tabulated} for the
tabulated results of the Case A+lapse scheme.

\section{Conclusions}
\label{sec:conclusions}

In this work, we have studied the performance of different pseudo-Newtonian approaches, which are motivated by 
state-of-the-art CCSN simulations, to approximate the neutron star oscillations in GR. 
We have derived four different perturbation schemes to study the radial and nonradial oscillation modes. 
Our analysis is built upon different combinations of hydrodynamics equations and gravitational potentials. 
We consider Newtonian hydrodynamics equations with or without a lapse-function correction. For the gravitational 
parts, we consider both the Newtonian potential and the Case A effective potential \citep{marek}. 
The four schemes include (see Table~\ref{tab:schemes}): purely Newtonian hydrodynamics (N), Newtonian hydrodynamics with lapse-function correction (N+lapse), Newtonian hydrodynamics with Case A potential (Case A), and Newtonian hydrodynamics with Case A potential and lapse-function correction (Case A+lapse). We focus in particular on the Case A+lapse scheme which is recently proposed by \citet{zha} in their CCSN simulations. 

For the radial fundamental $F$ mode, it is well known that the $F$ mode frequency would decrease toward zero as 
the background star model approaches the maximum-mass limit in GR, signifying the change of stability. We 
find that the N+lapse, Case A, and Case A+lapse schemes can capture this trend qualitatively, though the 
percentage differences between the mode frequencies computed by exact GR and these perturbation schemes 
increase significantly ($\sim \mathcal{O}(100\%)$) as the maximum-mass limit is approached. The purely 
Newtonian N scheme simply cannot capture this trend, which should already be expected. For star models with 
lower central density, we find that the N+lapse scheme can generally give a reasonable approximation to the
$F$ mode frequency to within about 1\% to 10\% levels, the values of which depend on the density and EOS model. 
On the other hand, the Case A+lapse scheme does not perform as good as the N+lapse scheme for the $F$ mode. 
The percentage differences between the $F$ mode frequencies computed by exact GR and the Case A+lapse scheme are typically a few tens of percent for star models with mass about $1.4 M_\odot$ for our EOS models. This level
of accuracy agrees with the study performed by \citet{zha} using numerical simulations (see the supplementary
material in their paper). 

For the nonradial fundamental $f$ mode, we find that the Case A+lapse scheme can approximate the mode frequency
very well to within a few percent level across the EOS models and central density ranges that we have considered. 
As the $f$-mode oscillations of a protoneutron star is expected to contribute strongly to the emitted 
gravitational wave signals, being able to determine the $f$ mode frequency of the protoneutron star
accurately in numerical simulations would be an important task. 
Our results would thus set a useful benchmark for numerical results obtained by CCSN simulations using 
different pseudo-Newtonian formulations. 
In particular, the original Case A formulation \citep{marek} would typically overestimate the $f$ mode frequency 
by about a factor of two. However, adding a lapse function correction to the hydrodynamics equations \citep{zha} can drastically improve the accuracy of the $f$ mode frequency.

Our finding that the Case A+lapse scheme can approximate the nonradial $f$ mode frequency in GR very well to 
within a few percent level, but generally overestimates the radial $F$ mode frequency by a few tens of percent 
might have some implications to the analysis of future simulations using this pseudo-Newtonian scheme. 
The postbounce oscillations of a protoneutron star can contain both radial and nonradial modes in a generic 
CCSN simulation. Nonlinear effects may excite modes with frequencies that are linear sums and differences 
of the linear $F$ and $f$ modes \citep{dimmelmeier,Passamonti:2007}. 
Furthermore, even if the star is initially dominated by the $F$ mode oscillations, the nonradial $f$ mode could 
also be excited strongly during the nonlinear evolution, if the frequencies of the two modes are close to each other.  As the $F$ mode frequency is not well approximated by the Case A+lapse scheme, any conclusion regarding these nonlinear effects seen in simulations should be treated with caution. For instance, a strong excitation of the $f$ mode, and hence an enhanced gravitational wave signals associated with it, due to its resonant coupling with the $F$ mode in a pseudo-Newtonian simulation may simply not occur in a corresponding GR modelling.

Finally, we end this paper by noting that the purely Newtonian scheme can also approximate the $f$ mode frequency quite well to within about 10\% accuracy even for the maximum-mass configuration in GR, if one chooses the central 
rest-mass density of the Newtonian star in such a way that it has the same central energy density as the GR 
counterpart via the EOS. Of course, the Newtonian and GR background stars are in general very different as expected, though their $f$ mode frequencies are close to each other.

\section*{Acknowledgements}

We thank Shuai Zha for helpful discussions and comments on the manuscript. This work is supported by a grant from the Research Grant Council of the Hong Kong Special Administrative Region, China (Project No. 14300320).

\section*{Data availability}

The data underlying this article are available in the article. No new data needed to be generated or analyzed.



\bibliographystyle{mnras}
\bibliography{paper} 




\appendix

\section{Boundary conditions for nonradial oscillations}
\label{sec:bc}

In this section, we present the derivation of the boundary conditions discussed in \Cref{sec:nonradial_bc}.
We first rewrite the perturbation equations in a form without intermediate variables by substituting 
equations~(\ref{n_iv1}) and (\ref{n_iv2}) into equations~(\ref{n_p1}), (\ref{n_p2}), (\ref{n_p3}), and 
(\ref{n_p4}). The rewritten perturbation equations are:
\begin{align}
\begin{split}
	\frac{dU}{dr}=-\left(\frac{2}{r}+\frac{d\Phi}{dr}+\frac{1}{\gamma P}\frac{dP}{dr} - \frac{A}{\alpha}\right)U 
		+ \left[\frac{\alpha l(l+1)}{\rho r^2 \omega^2}  \right . \\
		\left . -  \frac{1}{\alpha\Gamma_1 P}\right]\delta\tilde{P}  
	+ \frac{\alpha l(l+1)}{r^2 \omega^2}\delta\tilde{\Phi}  , 
\end{split} \label{bc_p1} \\
\begin{split}
	\frac{d\delta\tilde{P}}{dr}=\left(\frac{\rho\omega^2}{\alpha} - \frac{dP}{dr}A\right)U + \frac{1}{\Gamma_1 P}\frac{dP}{dr}\delta\tilde{P} - \rho\frac{d\delta\tilde{\Phi}}{dr}  ,
\end{split} \label{bc_p2} \\
\begin{split}
	\frac{d\Psi}{dr}=-\frac{2}{r}\Psi + \frac{l(l+1)}{r^2}\delta\tilde{\Phi} + 4\pi\frac{\rho}{\Gamma_1 P}\delta\tilde{P} - 4\pi\rho AU   ,
\end{split} \label{bc_p3} \\
\begin{split}
	\frac{d\delta\tilde{\Phi}}{dr}=\Psi,
\end{split}
\end{align}
where we have replaced $\frac{1}{\alpha}\frac{d\alpha}{dr}$ by $\frac{d\Phi}{dr}$. Note that when we turn off $\alpha$ in the N and Case A schemes, we should remove the $\frac{d\Phi}{dr}$ term in the above equations. 

At the center, we expand the perturbation variables according to 
\begin{align}
	&U=r^a\sum_n  A_n r^n , \\ 
	&\delta\tilde{P}=r^b \sum_n B_n r^n  , \\
	&\delta\tilde{\Phi}=r^c \sum_n C_n r^n  , \\
	&\Psi=r^{c-1} \sum_n (c+n)C_n r^n , 
\end{align}
where $a$, $b$, $c$, and the expansion coefficients $(A_n, B_n, C_n)$ are constants to be determined. 
Some background variables also scale with $r$ near the center and they are also expanded accordingly:  
\begin{align}
	&A = A'r \\
	&\frac{dP}{dr} = P''r \\
	&\frac{d\Phi}{dr} = -\frac{1}{\rho}\frac{dP}{dr} = -\frac{P''}{\rho}r.
\end{align}
We then substitute the above expansions into equations~(\ref{bc_p1}) to (\ref{bc_p3}). Keeping only the lowest order terms, we obtain
\begin{align}
	&br^{b-1}B_0=\frac{\rho\omega^2}{\alpha}r^aA_0 - \rho cr^{c-1}C_0 , \label{bc_expansion1} \\
	&(a+2)r^{a-1}A_0=\frac{\alpha l(l+1)}{\rho\omega^2}r^{b-2}B_0 + \frac{\alpha l(l+1)}{\omega^2}r^{c-2}C_0 , \label{bc_expansion2} \\
	&c(c+1)r^{c-2}C_0=l(l+1)r^{c-2}C_0 + 4\pi\frac{\rho}{\Gamma_1 P}r^bB_0 - 4\pi\rho A'r^{a+1}A_0 .
\end{align}
We demand that $a$, $b$ and $c$ be independent of background variables, and that all of them can be determined simultaneously. By considering all possible cases and by elimination (see Appendix A in \citet{WS2020} for more detail), it can be shown that $a+1=b=c$, and $c=l$. Therefore, we have 
\begin{align}
	&U=r^{l-1}A_0 , \label{bc_cbc1}\\
	&\delta\tilde{P}=r^lB_0 , \label{bc_cbc2}\\
	&\delta\tilde{\Phi}=r^lC_0 , \label{bc_cbc3}\\
	&\Psi=lr^{l-1}C_0. \label{bc_cbc4}
\end{align}
Substituting these expansions back into equation~(\ref{bc_expansion1}), we obtain  
\begin{equation}
	A_0=\frac{\alpha l}{\rho\omega^2}(B_0+\rho C_0). \label{bc_cbc5}
\end{equation}
Equations~(\ref{bc_cbc1}) to (\ref{bc_cbc5}) are the central boundary conditions. For the N and Case A schemes where the lapse function $\alpha$ is turned off, the central boundary conditions are the same except that the 
term $\alpha$ in equation~(\ref{bc_cbc5}) becomes unity.

For the surface boundary conditions, the derivation is simpler. The first boundary condition comes from the requirement that the Lagrangian perturbation of pressure should vanish at the surface:
\begin{equation}
	\Delta P=\delta P + \vec{\xi}\cdot\nabla P=\delta P+\frac{dP}{dr}U=0,
\end{equation}
where $\Delta P$ denotes the Lagrangian perturbation of pressure and $\delta P$ is the Eulerian perturbation of pressure. The second surface boundary condition is obtained by requiring $\delta\tilde{\Phi}$ (or $\delta\Phi$) and its derivative be continuous across the surface. In the following, we use the subscript $i$ or $o$ on a 
variable to denote whether the variable is evaluated inside or outside the star surface, respectively.  
We first evaluate the perturbed Poisson equation outside the star, $\nabla^2\delta\Phi_o=0$, by expanding 
$\delta\Phi_o$ in spherical harmonics (i.e., $\delta\Phi_0 = \delta{\tilde \Phi}_o Y_{lm}$). 
The perturbed Poisson equation becomes
\begin{equation}
	r^2 \frac{d^2}{dr^2}\delta\tilde{\Phi}_o + 2r\frac{d}{dr}\delta\tilde{\Phi}_o - l(l+1)\delta\Phi_o = 0.
\end{equation}
This equation can be solved exactly and we obtain
\begin{equation}
	\delta\tilde{\Phi}_o = \frac{C}{r^{l+1}}  . \label{bc_sbc0}
\end{equation}
To extract the boundary condition, we differentiate \cref{bc_sbc0} and eliminate $C$ using \cref{bc_sbc0} again. Since $\delta\tilde{\Phi}$ and its first derivative are continuous, we obtain
\begin{equation}
	\frac{d\delta\tilde{\Phi}_i}{dr}=-\frac{l+1}{R}\delta\tilde{\Phi}_i.
\end{equation}
Dropping the subscripts and replacing $d\delta\tilde{\Phi}/dr$ by the variable $\Psi$, we obtain the second 
surface boundary condition:
\begin{equation}
	\Psi = -\frac{l+1}{R}\delta\tilde{\Phi}.
\end{equation}


\section{Tabulated results}
\label{sec:tabulated}

In this Appendix, we tabulate the results for the Case A+lapse scheme using EOS A, B, C and F. In each of the following tables, for a given central energy density 
$e_c$, we present our numerical results for the mass $M$ and radius $R$ of the background
star; the frequencies of the radial fundamental $F$ mode, its first overtone $H_1$, the nonradial quadrupolar $f$ mode, and the first pressure $p_1$ mode. 
The numerical values inside the parentheses represent the percentage differences between
our numerical results and the exact GR results given in \citep{andersson,kokkotas}.

\begin{table*}
	\centering
	\caption{Tabulated results for EOS A.}
	\label{tab:r_eosA}
	\begin{tabular}{ccccccc}
		\hline
		$e_c (10^{15}\ {\rm gcm}^{-3})$ & $M (M_\odot)$ & $R ({\rm km})$ 
		 & $F ({\rm Hz})$ & $H_1 ({\rm Hz})$ & $f ({\rm Hz})$ & $p_1 ({\rm Hz})$\\
		\hline
		1.259 & 1.05\ (0.1) & 10.38\ (4.9) & 4382\ (19.4) & 8981\ (11.7) & 2330\ (5.8) & 8564\ (13.5) \\
		1.698 & 1.32\ (0.4) & 10.36\ (7.2) & 4147\ (28.4) & 10084\ (17.7) & 2575\ (5.2) & 9446\ (21.0)  \\
		1.995 & 1.45\ (0.1) & 10.28\ (8.5) & 3958\ (34.6) & 10289\ (21.6) & 2705\ (4.9) & 9750\ (24.7) \\
		2.344 & 1.54\ (0.2) & 10.16\ (9.8) & 3693\ (40.9) & 10342\ (24.8) & 2826\ (4.5) & 10011\ (28.0) \\
		3.000 & 1.63\ (0.8) & 9.94\ (12.0) & 3077\ (56.2) & 10267\ (28.7) & 2988\ (3.5) & 10362\ (32.5) \\
		3.980 & 1.68\ (1.8) & 9.64\ (14.5) & 1765\ (167.4) & 9995\ (31.0) & 3145\ (1.8) & 10696\ (36.5) \\
		\hline
	\end{tabular}
\end{table*}

\begin{table*}
	\centering
	\caption{Tabulated results for EOS B.}
	\label{tab:r_eosB}
	\begin{tabular}{ccccccc}
		\hline
		$e_c (10^{15}\ {\rm gcm}^{-3})$ & $M (M_\odot)$ & $R ({\rm km})$ 
		 & $F ({\rm Hz})$ & $H_1 ({\rm Hz})$ & $f ({\rm Hz})$ & $p_1 ({\rm Hz})$\\
		\hline
		1.995 & 0.97\ (0.0) & 9.30\ (6.1) & 4283\ (19.6) & 9826\ (13.6) & 2761\ (3.8) & 8514\ (14.0) \\
		3.000 & 1.25\ (0.2) & 8.93\ (9.6) & 4483\ (34.2) & 10730\ (21.8) & 3242\ (4.2) & 10078\ (21.2) \\
		3.388 & 1.31\ (0.4) & 8.81\ (10.7) & 4332\ (39.7) & 10966\ (23.8) & 3366\ (4.0) & 10517\ (23.5)  \\
		5.012 & 1.43\ (1.5) & 8.39\ (14.7) & 3155\ (73.3) & 11429\ (28.3) & 3688\ (2.5) & 11684\ (30.4) \\
		\hline
	\end{tabular}
\end{table*}

\begin{table*}
	\centering
	\caption{Tabulated results for EOS C.}
	\label{tab:r_eosC}
	\begin{tabular}{ccccccc}
		\hline
		$e_c (10^{15}\ {\rm gcm}^{-3})$ & $M (M_\odot)$ & $R ({\rm km})$ 
		 & $F ({\rm Hz})$ & $H_1 ({\rm Hz})$ & $f ({\rm Hz})$ & $p_1 ({\rm Hz})$\\
		\hline
		1.000 & 1.33\ (0.6) & 12.75\ (6.1) & 3100\ (19.7) & 6960\ (13.5) & 1959\ (3.9) & 6301\ (14.5) \\
		1.122 & 1.44\ (0.0) & 12.66\ (7.0) & 3134\ (22.4) & 7191\ (15.1) & 2047\ (3.7) & 6570\ (16.0) \\
		1.413 & 1.62\ (0.0) & 12.42\ (8.6) & 3056\ (29.5) & 7501\ (18.5) & 2220\ (3.5) & 7088\ (19.6) \\
		1.778 & 1.75\ (0.2) & 12.15\ (10.3) & 2796\ (38.4) & 7719\ (21.8) & 2375\ (3.1) & 7542\ (23.1) \\
		1.995 & 1.80\ (0.4) & 11.99\ (11.2) & 2604\ (44.6) & 7819\ (23.3) & 2449\ (2.8) & 7751\ (24.8) \\
		3.000 & 1.90\ (2.6) & 11.37\ (14.3) & 1145\,(281.8) & 7984\ (28.2) & 2694\ (1.4) & 8410\ (30.8) \\
		\hline
	\end{tabular}
\end{table*}

\begin{table*}
	\centering
	\caption{Tabulated results for EOS F.}
	\label{tab:r_eosF}
	\begin{tabular}{ccccccc}
		\hline
		$e_c (10^{15}\ {\rm gcm}^{-3})$ & $M (M_\odot)$ & $R ({\rm km})$ 
		 & $F ({\rm Hz})$ & $H_1 ({\rm Hz})$ & $f ({\rm Hz})$ & $p_1 ({\rm Hz})$\\
		\hline
		1.122 & 1.04\ (0.5) & 11.39\ (4.6) & 3217\ (17.0) & 7318\ (11.4) & 2063\ (4.2) & 6793\ (12.5)  \\
		1.585 & 1.23\ (0.4) & 11.13\ (6.4) & 2928\ (24.1) & 7594\ (14.7) & 2306\ (3.4) & 7390\ (16.2) \\
		2.239 & 1.34\ (0.8) & 10.75\ (8.3) & 2460\ (33.7) & 7973\ (17.9) & 2553\ (3.0) & 7888\ (19.8) \\
		3.162 & 1.44\ (1.7) & 10.13\ (11.5) & 2328\ (42.8) & 9035\ (22.1) & 2897\ (1.3) & 8526\ (24.4) \\
		3.981 & 1.48\ (2.4) & 9.69\ (14.1) & 2176\ (53.2) & 9503\ (26.0) & 3143\ (0.1) & 9025\ (27.4) \\
		5.012 & 1.51\ (3.3) & 9.29\ (16.7) & 1341\ (191.5) & 9528\ (28.4) & 3356\ (1.4) & 9514\ (29.5) \\
		\hline
	\end{tabular}
\end{table*}


\bsp	
\label{lastpage}
\end{document}